\documentstyle[preprint,aps,prl,epsf]{revtex}
\def\upket{ \vert \,{\scriptstyle\uparrow}\,\rangle}
\def\downket{ \vert \,{\scriptstyle\downarrow}\,\rangle}

\input{epsf}
\input epsf.sty
\begin{document}
\draft
\title{
Anisotropic Transport of Quantum Hall Meron-Pair Excitations  
}
\author{Kyungsun Moon and Kieran Mullen}
\address{
Department of Physics and Astronomy, University of Oklahoma, Norman,
OK~~73019\\
}
\date{\today}
\maketitle

{\tightenlines
\begin{abstract}
Double-layer quantum Hall systems at total filling factor $\nu_T=1$
can exhibit a commensurate-incommensurate phase transition
driven by a magnetic field $B_{\parallel}$ oriented parallel to the layers.
Within the commensurate phase, the lowest charge excitations are believed
to be {\it linearly-confined Meron pairs}, which are energetically
favored to align with $B_{\parallel}$.
In order to investigate this interesting object, we propose a gated 
double-layer Hall bar experiment in which $B_{\parallel}$ 
can be rotated with respect to the direction of a constriction.  
We demonstrate the strong angle-dependent transport
due to the anisotropic nature of linearly-confined Meron pairs 
and discuss how it would be manifested in experiment.
\end{abstract}
}


\narrowtext

The quantum Hall effect has served as a canonical example of a 
strongly correlated, two-dimensional 
electron system. Since the strong perpendicular
magnetic field $B_{\perp}$ quenches the kinetic energy of the electrons,  
the electron-electron correlations become crucial.
These correlations can manifest themselves in 
a variety of 
{\em topological charges}:
 patterns or textures in the spin or
isospin of the 2D electrons. 
In the case of a single-layer quantum Hall system,
Sondhi {\em et al.} have shown that even when the Zeeman gap vanishes, the
quantum Hall effect survives at some filling factors including $\nu=1$
\cite{Sondhi,DBPRBI}.  
They argued that in this case the lowest energy charged excitations
are Skyrmions, topological excitations consisting of
dimples in the electron spin distribution which 
involve multiple spin-flips\cite{Sondhi,DBPRBI,MacdFertig,WuandJain}.
Subsequently S.E. Barrett {\em et al.} performed NMR Knight shift 
measurements and clearly demonstrated the existence of multiple spin-flip 
excitations which may be a manifestation of these Skyrmionic 
excitations\cite{Barrett,Eisenstein,Aifer}.

The closely related 
double-layer quantum Hall systems (DLQHS) have also drawn much theoretical 
and experimental 
attention\cite{DBPRBI,DBPRL,DBPRBII,GirvinMacdBook,Ezawa,Sheena,Shayegan}.
When the distance $d$ between the layers is comparable to the mean intralayer
particle spacing, strong interlayer correlations induce a novel
manybody quantum Hall effect at $\nu_T=1/m$ ($m$ is an odd integer), 
where $\nu_T$ is the total filling factor.
The layer degrees of freedom ({\em i.e.} upper or lower)
can be viewed as a spin ${1\over 2}$ {\em isospin} variable, and the description
of
DLQHS can be mapped onto itinerant quantum ferromagnets
\cite{DBPRBI,DBPRL,DBPRBII,TLHo}. We 
emphasize that the real spins are fully spin-polarized along $B_{\perp}$
due to the Zeeman gap and the excitations consist of variations in
the isospin\cite{Kmoon}.
It has been argued that for the experimentally relevant set of parameters, 
linearly-confined Meron pairs (LMP) are the lowest energy
 charged excitations.   They too 
are genuine topological excitations made 
from textures in the isospin distribution\cite{DBPRBII,Nick} as shown in
Fig.(\ref{lmp}). 
In contrast to Skyrmions, 
these excitations cannot be seen via NMR Knight shift measurements because
they do not couple to the nuclear magnetic moment.
Such excitations are notoriously difficult to observe
and have been seen only indirectly via transport measurements\cite{Sheena}.

In this paper we propose a  novel method of investigating topological 
excitations by forcing them to pass through a narrow channel.  
As the constriction is approached, details of the excitations can be measured.
This method
can be used on systems in which it is hard to couple with the relevant
isospin ({\em e.g.}  valley-Skyrmions in silicon 2D electronic 
systems\cite{valley}).
Here we concentrate on the LMP: 
by noticing that the orientation of the LMP prefers to follow the direction 
of the magnetic field $B_{\parallel}$ applied parallel to the layers,   
we demonstrate the strong transport anisotropy depending on the
relative angle between the gate and $B_{\parallel}$.
The method can be 
generalized to other classes of topological objects\cite{tobe}.

Following the magnetic analogy, we map the layer degrees of freedom 
onto a $S={1\over 2}$ isospin variable,
where an electron in the upper-layer  has an isospin 
$\upket$ and one in the lower-layer has an isospin
$\downket$\cite{DBPRBI}. 
A local charge imbalance between the two layers corresponds to 
$\langle S_z({\bf r}) \rangle \ne 0$ in that region.
Such charge fluctuations between the layers have an energy gap at long
wavelengths and are suppressed.
Thus while the
true spin of the electron points in the $\hat z$ direction, 
the isospin is forced to lie in the ${\hat x}-{\hat y}$ plane in 
isospin space, so the system is equivalent to a quantum $XY$-ferromagnet.
Using the spin texture state ansatz    
$\vert\theta({\bf r})\rangle=
\left(\upket+ e^{i \theta({\bf r})}\downket\right)/2$,
we obtain the following energy functional $E[\theta]$ 
\begin{equation}
E[\theta]=\int d^2r~\Biggl\{ \frac{1}{2}~\rho_s ({\bf\nabla}\theta)^2 -
{t\over 2\pi\ell_B^2}~\cos{\theta ({\bf r})}\Biggr\} 
\end{equation}
where $\theta({\bf r})$ represents the isospin orientation in the plane, 
$\rho_s$ is the isospin
stiffness, $t$ stands for 
an interlayer tunneling amplitude, and $\ell_B=(\hbar c/|e|B_{\perp})^{1/2}$.

The first term indicates that the isospins prefer to be parallel to 
each other, which minimizes the overlap of the electrons due to the Pauli
exclusion principle and so minimizes
 the Coulomb exchange energy. The second term 
reflects the fact that the tunneling term lowers the energy of
electrons in a symmetric
superposition of the two layer states, which corresponds to 
$\langle S_x \rangle= {1\over 2}$.  This favors orienting the isospin along  
the ${\hat x}$-direction in isospin space.
Hence the ground state of this Hamiltonian will be the Slater determinant of
the completely filled symmetric state.

The presence of an in-plane
magnetic field $B_{\parallel}$ makes the physics of
DLQHS much more intriguing. Suppose $B_{\parallel}$ is parallel to the 
${\hat y}$-coordinate.
Choosing the gauge potential 
${\bf A}=(0,B_{\perp} x,-B_{\parallel} x)$,  we see that
the tunneling matrix element from one layer to the other will pick up
an Aharanov-Bohm phase $e^{i Q x}$ with $Q=dB_{\parallel}/\ell_B^2 B_{\perp}$, 
and that
this phase rotates or ``tumbles'' as we move in $x$.  
The imposition of $B_{\parallel}$ leaves the orbital degrees of
freedom intact, and
we obtain the following energy functional  for the ansatz
given above:
\begin{equation}
E[\theta]=\int d^2r~\Biggl\{ \frac{1}{2}~\rho_s ({\bf\nabla}\theta)^2 -
{\bf h}({\bf r})\cdot{\bf m}({\bf r})\Biggr\} 
\end{equation}
where the fictitious magnetic field ${\bf h}({\bf r})={t\over 2\pi\ell_B^2} 
(\cos Qx,\sin Qx)$ tumbles along the ${\hat x}$-coordinate with 
a period $2\pi/Q$, which couples to the isospin  
${\bf m}({\bf r})=\left(\cos\theta({\bf r}),\sin\theta({\bf r})\right)$.
This is the well-known Pokrovsky-Talapov model, which 
exhibits a highly collective commensurate-incommensurate 
transition\cite{bak}.
For small $Q$ and/or small $\rho_s$, the phase tracks the tumbling field,
so that 
$\theta({\bf r})= Qx$. As $B_{\parallel}$ increases, 
the local field tumbles too rapidly and a continuous
phase transition to an incommensurate state with broken
translation symmetry occurs\cite{DBPRBII,bak}.  
The ground state energy of the commensurate state is given by 
\begin{equation}
\frac {E_0[Q]}{A}={1\over 2}\rho_s Q^2 -{t\over 2\pi\ell_B^2} .
\end{equation}
For the clarity of further discussion, we introduce a phase field 
$\phi({\bf r})=\theta ({\bf r}) - Qx$,
which represents an isospin orientation measured with respect to the 
direction of ${\bf h}({\bf r})$.
For the commensurate state, the isospin is parallel to ${\bf h}({\bf r})$ 
yielding $\phi({\bf r})=0$.
The energy functional for the generic excited states can be written 
in terms of the $\phi({\bf r})$-field as follows
\begin{equation}
E_A[\phi]= \rho_s Q L_y \Delta\phi  
+\int d^2r~\Biggl\{ \frac{1}{2}~\rho_s ({\bf\nabla}\phi)^2 
-{t\over 2\pi\ell_B^2}~(1-\cos{\phi ({\bf r})})\Biggr\}  
\label{excite}
\end{equation}
where $L_y$ is the system size along the ${\hat y}$-direction
and $\Delta\phi\equiv\phi(x=\infty)-\phi(x=-\infty)$, the number of full
rotations the isospin makes with respect to the tumbling field
${\bf h}({\bf r})$, 
is a {\em topological} charge\cite{coleman}.   

These rotations or phase
slips with $\Delta\phi=\pm 2\pi$
occur over a finite width and can be viewed as domain walls\cite{bak}. 
If the domain-wall string soliton (DWS)  has finite length, its endpoints
will be localized excitations in the isospin texture called Merons. The
domain wall serves to link the Meron pair so that they are linearly
confined, as shown in 
fig.(\ref{lmp}).  It has been argued that the LMP is 
the lowest energy {\em charged} excitation of the DLQHS 
within the commensurate phase\cite{DBPRBI,DBPRBII,Kmoon}.  

Suppose we have a LMP extending over the system size and making an angle 
$\alpha$ with $B_{\parallel}$.
Since the second term of Eq.(\ref{excite}) is invariant under spatial 
rotations (that is, rotations in ${\bf r}$), 
dependence of the LMP activation energy on its orientation in the plane
comes entirely
from the first term, which depends on $L_y$-the projected length 
to the ${\hat y}$-direction.
The first term can be viewed as a chemical potential for the LMP 
with $\mu_D=-2\pi\rho_s Q L_y$. 
We notice that for the non-topological excitations with $\Delta\phi=0$,
the activation energy has {\em no} angle-dependence.
For the sake of simplicity, we can take the LMP to be aligned along the 
${\hat y}$-axis.
The analytical solution for the profile of the domain wall
is well-known and given by  
$\phi(x)=-4\tan^{-1}[\exp((2\pi \rho_s/t)^{1/2}\, x/\ell_B)]$\cite{bak}.
Based on this solution, the energy $T_0$ of the LMP per unit length is  
given to be $(4t/\pi\ell_B)(2\pi\rho_s/t)^{1/2}$.
Hence the string tension is given by
\begin{equation}
T[\alpha,B_{\parallel}]=T_0 \left\{\left(1-{B_{\parallel}\over B_{\parallel}^c}\right)
+ {B_{\parallel}\over B_{\parallel}^c} \left(1-\cos\alpha\right)\right\} 
\end {equation}
where $B_{\parallel}^c=B_{\perp}(4\ell_B/\pi d)(t/2\pi\rho_s)^{1/2}$.
When $B_{\parallel}>B_{\parallel}^c$, it is energetically favorable to create
 DWSs of infinite length (in other words, the LMP's become unbound)
making a phase transition to an incommensurate 
phase\cite{DBPRL,DBPRBII,Sheena}.
Within the commensurate phase $B_{\parallel}<B_{\parallel}^c$, the activation energy 
of the DWS increases linearly with the length.

Since the merons carry charge $\pm {1 \over 2}e$ depending on the vorticity
and core-spin configurations\cite{DBPRBI},
one can construct a finite-energy charged excitation 
by attaching two Merons with the same charge and opposite vorticity.
The activation energy $E_{\rm LMP}$ of the LMP with the length $R$ and 
the relative angle $\alpha$ with respect to $B_{\parallel}$ can be determined
by balancing the
Coulomb repulsion and the linear string tension 
\begin{equation}
E_{\rm LMP}=T(\alpha,B_{\parallel})R +\frac {e^2}{4\epsilon R} + 2E_{mc}
\label{balance}
\end {equation}
where $\epsilon$ is the dielectric constant and 
$E_{mc}$ represents the Meron core energy obtained by integrating out 
the short-distance degrees of freedom.
Eq.(\ref{balance}) is optimized when the LMP is oriented along 
$B_{\parallel}$, that is, $\alpha=0$.
The equilibrium distance $R_c$ is given by   
\begin{equation}
R_c=\left(\frac {e^2}{4\epsilon\,T(\alpha=0,Q)}\right)^{1/2}\;\propto\; 
(1-\rho)^{-1/2}
\end{equation}
where $\rho=B_{\parallel}/B_{\parallel}^c$ is a magnetic field measured 
in units of critical value $B_{\parallel}^c$.  
It is  amusing to note that the rotations of $B_{\parallel}$ 
can be used as a knob to orient the LMP.
As $B_{\parallel}$ increases, the string tension decreases
as $(1-\rho)$ and the length of the LMP increases with $(1-\rho)^{-1/2}$.
At finite temperature, one needs to take into account the effect of 
thermal fluctuations which can distort this object via stretching {\em or} 
rotation.
The energy cost $\Delta E(\alpha,R,B_{\parallel})$ of small fluctuations 
over the optimal solution of the LMP is given by  
\begin{equation}
\Delta E[\alpha,R,B_{\parallel}]\cong {1\over 2} \kappa_R 
\left(1-\frac {R}{R_c}\right)^2 + {1\over 2} \kappa_{\alpha}\alpha^2
\end{equation}
where the spring constants $\kappa_R=e^2/(2\epsilon R_c)\propto (1-\rho)^{1/2}$ 
and $\kappa_\alpha=\rho T_0 R_c\propto \rho/(1-\rho)^{1/2}$.
In order to detect this interesting object,
we propose a gated Hall bar experiment where the relative
orientation of the constriction with respect to $B_{\parallel}$  
can be varied.  
In Fig.(2), we have shown a quantum Hall bar which is gated
in the middle by putting
metallic gates in both layers. The constriction has a channel width $W$, 
which can be varied by adjusting the gate voltage.
For simplicity, the channel  is assumed to have a 
`hard wall' which prevents the transport of charge carriers.
The $\nu_T=1$ state can be considered as a vacuum of the LMP.
The perpendicular magnetic field $B_{\perp}$ is applied so that 
the total filling factor $\nu_T$ of the system 
is slightly away from $1$.  
Since the lowest charge excitations are argued to be the LMP,  
the ground state of the system will have LMP's.
Since the LMP prefers to be parallel to $B_{\parallel}$,
the transport through a constriction will have a strong 
dependence on the relative 
angle $\psi$ between $B_{\parallel}$ and the constriction.
The transport probability $T_{\rm tr}(B_{\parallel},\psi)$ of the LMP
passing through a narrow constriction is given by
\begin{equation}
T_{\rm tr}(B_{\parallel},\psi)\sim |{\cal T}|^2 \int_0^{2\pi} d\alpha 
\int_0^{W_e\over |\cos(\psi-\alpha)|}dR\,
(W_e-R|\cos(\psi-\alpha)|) 
\,e^{-\beta\,\Delta E(\alpha,R,B_{\parallel})}
\end{equation}
where $\psi$ is a relative angle between the gate and $B_{\parallel}$, 
$\beta$ is inverse temperature,    
and $|{\cal T}|^2$ is a transmission coefficient.
Since the channel width $W$ should be larger than the Meron core size
$R_{\rm mc}$ which is estimated to be about $2 \ell_B$\cite{YangMacd,Brey},
the effective channel width $W_e$ is set to be $W-2R_{\rm mc}$.
We assume that $|{\cal T}|^2$ has {\em no} angle dependence\cite{comment1}. 
As $B_{\parallel}$ approaches $B_{\parallel}^c$, $\kappa_R$
vanishes as $(1-\rho)^{1/2}$ and $\kappa_\alpha$ diverges as 
$\rho/(1-\rho)^{1/2}$. In this limit, $T_{\rm tr}(B_{\parallel},\psi)$ can be
obtained analytically 
\begin{equation}
T_{\rm tr}(B_{\parallel},\psi)\sim |{\cal T}|^2 W_e^2 
\frac {(1-\rho)^{1/4}}{\cos\psi}
\frac {(k_B T)^{1/2}}{(T_0 e^2/4\epsilon)^{1/4}}.
\end{equation}
Note the strong angle dependence of $T_{\rm tr}(B_{\parallel},\psi)\propto  
1/\cos\psi$.
At $\psi=0$, the LMP tends to be parallel to the constriction. 
Since the LMP which is larger than the narrow channel 
can not easily pass through it,  
the transport probability rapidly decreases as $B_{\parallel}$
gets to $B_{\parallel}^c$, where the length of the LMP becomes very large.
If we rotate the field by $\pi/2$,
 the LMP prefer to be oriented perpendicular 
to the constriction, which will strongly enhance the transport probability. 
We define ${\cal A}(B_{\parallel})$ to be the ratio
of $T_{\rm tr}(B_{\parallel},\psi=0)$ to $T_{\rm tr}(B_{\parallel},\psi=\pi/2)$
\begin{equation}
{\cal A}(B_{\parallel})\equiv \frac {T_{\rm tr}(B_{\parallel},\psi=0)}
{T_{\rm tr}(B_{\parallel},\psi=\pi/2)}
\end{equation}
which measures the transport anisotropy. 
Based on the experiment by S.Q. Murphy {\em et al.}\cite{Sheena},
we have chosen the following set of parameters:
the Coulomb energy $e^2/\epsilon\ell_B\cong 130 {\rm K}$
and $t\cong 0.5 K$.
The isospin stiffness $\rho_s$ is estimated to be about $0.5 K$
and the string tension $T_0$ is about $1.6 K$\cite{DBPRBI,Kmoon}.
At $B_{\parallel}/B_{\parallel}^c=0.9$, the length of the LMP is estimated
to be
about $15\ell_B\sim 1400 \AA$.
We have chosen two values of $W_e$ to be $5\ell_B, 8\ell_B$ 
and the temperature is set to be $300 mK$.

Fig.(3) shows ${\cal A}(B_{\parallel})$ as a function of $\rho$.
We notice that at $B_{\parallel}=0$, transport is isotropic as expected, since 
the anisotropy is due to a finite $B_{\parallel}$.
As $B_{\parallel}$ increases and approaches to $B_{\parallel}^c$, 
the anisotropy drastically increases, which we believe can be a clear 
signature to identify the LMP.
This anisotropy can only be seen below a certain
temperature $T_{\rm KT}$.
In the absence of tunneling, we expect the Kosterlitz-Thouless
phase transition to occur at $T_{KT}$\cite{WenandZee,DBPRBI}.  
Above $T_{KT}$, there will be many
free Merons which carry a charge $\pm {1\over 2}e$.
Finite tunneling converts the KT-trasition into a cross-over
due to the explicitly broken $U(1)$-symmetry.
Hence our picture of the linearly-confined Meron pairs as  
the lowest charged excitations holds below $T_{\rm KT}$.
The transition temperature $T_{\rm KT}$ is estimated to be about 
${\pi\over 2}\rho_s\sim 0.6 K$\cite{comment2}.  We notice that 
the temperature dependence of ${\cal A}(B_{\parallel})$ is weak well
below the transition.

To summarize, we propose that topological charges in 2D electronic systems
can be probed by a gate geometry.  This is especially important if the
excitation is based upon an isospin that couples poorly to most experimentally
controllable parameters.  We 
 have shown that for  the case of linearly-confined Meron pairs there is
a strong transport anisotropy due to its topological nature.
In order to detect this fascinating object,
we propose a transport experiment through a narrow constriction with a variable
angle between the constriction and the parallel magnetic 
field $B_{\parallel}$.
We have clearly demonstrated that the transport has a strong 
angular dependence as $B_{\parallel}$ gets near 
to the critical value $B_{\parallel}^c$.   In other cases parameters such
as the size, energy and stiffness of the topological excitation might be
probed.

\acknowledgements{
It is our pleasure to acknowledge useful conversations with S.M. Girvin,
A.H. MacDonald, B.A. Mason, S.Q. Murphy, and J.P. Eisenstein. 
The work was supported by NSF DMR-9502555. 
K. Moon wants to acknowledge the Aspen center for physics where part 
of this work has been performed.
}

\begin{figure}
\caption{Isospin configurations of a finite-length Meron pair excitations: 
The arrows represent the isospin orientations. Merons with the opposite
vorticy and the same charge ${1\over 2}e$ are located at both ends 
of domain-wall. Inset shows the solution profile of 
$\phi(x)$ along a line passing between the Meron pair. \label{lmp}
}
\label{fig1}
\end{figure}

\begin{figure}
\caption{Schematic diagram of a gated Hall bar: 
the domain-wall string soliton is a linearly confined Meron pair.
$B_{\parallel}$ is the magnetic field applied parallel to the layer and
$B_{\perp}$ is a strong perpendicular magnetic field. 
$\psi$ is the relative angle between 
$B_{\parallel}$ and the constriction.  
Depending on $\psi$ and $B_{\parallel}$, the Meron pair will 
either easily pass 
through the gated region or be blocked.
}
\label{fig2}
\end{figure}

\begin{figure}
\caption{Transport anisotropy of linearly-confined Meron pair excitations:
The effective width of constriction $W_e$ is chosen to be $5\ell_B, 
8\ell_B$.
${\cal A}(B_{\parallel})$ is plotted as a function of   
$\rho=B_{\parallel}/B_{\parallel}^c$ at $T\sim 300 mK$.
}
\label{fig3}
\end{figure}


\begin{references}
\bibitem{comment2} The observed collapse of the activation gap at around 
$0.5 K$ in spite of much larger activation energy on the order of $10 K$
indicates that the charged excitation indeed has a 
manybody origin\cite{Sheena}.
\end{references}
\end{document}